# Subwavelength plasmonic chain for magneto-optics enhancement

D. G. Baranov,[1,2] A. P. Vinogradov,[1,2,3] and A. A. Lisyansky[4]

[1]*Moscow Institute of Physics and Technology, 9 Institutskiy per., Dolgoprudniy 141700, Russia*
[2]*All-Russia Research Institute of Automatics, 22 Sushchevskaya, Moscow 127055, Russia*
[3]*Institute for Theoretical and Applied Electromagnetics, 13 Izhorskaya, Moscow 125412, Russia*
[4]*Department of Physics, Queens College of the City University of New York, Queens, NY 11367, USA*

We study the enhancement of the magneto-optical effect in subdiffraction plasmonic chains. We show that in a periodic chain of the plasmonic nanoparticles embedded in a magneto-optical medium, propagation of a guided mode is accompanied by rotation of electrical dipoles (the Faraday effect). The angle of rotation per 1 μm is two orders of magnitude greater than that in the same bulk magneto-optical medium. We also demonstrate that the effect of Ohmic losses can be significantly reduced by using a gain-assisted chain composed of active core-shell nanoparticles (spasers). The dipole mode in such an array of magneto-optical spasers exhibits high values of the Faraday rotation and propagation length.

## I. INTRODUCTION

The magneto-optical (MO) effect is widely used to manipulate light. Time-reversal symmetry breaking due to the presence of a magnetic field in MO materials offers various opportunities for effective polarization conversion and optical isolation.[1] However, the MO response of natural media is relatively weak so that enhancements of MO properties is highly desirable. Recently, a number of approaches to enhancing MO effect have been proposed (for a review see, e.g., Ref. [2]). One of these incorporates layered systems[3-9] for which a rigorous eigenmodes analysis can be performed. In these structures, at a certain frequency, multiple refraction of light from interfaces between magnetic and non-magnetic layers may result in the Fabry-Perot resonance and an enhancement of the Faraday rotation. The enhancement is due to the fast growth of the phase, which changes by $\pi$ when the frequency passes through the resonance. Thus, instead of a small bulk value of the Faraday angle, one can obtain the value of order of $\pi/2$.[2,10]

MO effects can also be enhanced in metallic systems. The effect of the plasmon resonance of charge carries on the MO response was first analyzed in Ref. [11] in which strong Kerr effect enhancement in bulk metal near the plasma frequency was reported (see also Refs. [12,13]). Strong coupling between travelling surface plasmons in metals and MO effects was also demonstrated using the total internal reflection technique.[14]

Localized surface plasmon excitation in MO composites containing metal nanoparticles (NPs) also promise to strongly enhance the MO response of these materials.[15-20] A different type of the MO enhancement attributed to the phenomenon of extraordinary optical transmittance was observed in systems with magnetic metallic gratings[21, 22] and perforated periodic structures.[23-25]

In this paper, we propose another approach to resonance MO enhancement. This approach incorporates properties of so-called *subdiffraction chains* (SDCs).[26] We investigate the enhancement of the Faraday rotation in linear periodic SDCs of plasmonic NPs embedded in a MO host medium. We show that in this system the Faraday effect can be strongly enhanced and discuss how to compensate for Ohmic losses in metallic NPs.

Periodic linear chains of near-field coupled plasmonic NPs have been studied extensively since the original work of Quinten et al.[26] It was shown that a one-dimensional (1D) NP array supports guided modes due to near-field interactions between adjacent NPs. At first, only quasistatic interactions with the nearest neighbors were considered.[27, 28] Then, the effects of retardation that have significant impact on the dispersion characteristics of long wavelength guided modes were taken into account.[29] Modes of finite 1D,[30] infinite two-[31] and three-dimensional[32] as well as disordered arrays[33] of NPs were also investigated.

Intriguing properties of subdiffraction plasmonic chains composed of magnetized NPs have been studied recently. In particular, the combination of MO activity and geometrical chirality of a chain, made of twisted ellipsoidal NPs, have been shown to lead to a one-way subdiffraction waveguide.[34, 35] However, the Faraday effect has not been investigated in these systems. In Ref. [36], the Faraday rotation in a two-dimensional lattice of $Fe_3O_4$ NPs was studied experimentally. Unfortunately, the theoretical analysis presented in the paper is limited to a pair of coupled magnetite NPs. A more detailed theoretical investigation of magnetic NPs clusters was presented in Ref. [37], in which the Faraday rotation and circular dichroism were studied for the configurations of dimers, helices, and a random NP gas. In all these systems, an enhancement of the Faraday rotation due to plasmonic resonance is predicted.

The paper is organized as follows: In Section II, we derive an analytical expression for the polarizability of an isotropic spherical NP embedded into a gyrotropic medium. In Section III, we consider a 1D periodic array of silver NPs in a gyrotropic host medium. In Section IV, we analyze guided modes of a 1D gain-assisted MO SDC. The results are summarized in the Conclusion.

## II. NANOPARTICLE EMBEDDED INTO A MAGNETO-OPTICAL MEDIUM

We study a periodic chain of spherical metallic NPs of radius $R$ and inter-particle separation $L$ embedded into a MO host medium. This structure is shown schematically in Fig. 1. We assume that the chain is aligned in the direction of the MO medium magnetization vector, which is parallel to the $z$-axis. The permittivity tensor has the form:

$$\hat{\varepsilon} = \begin{pmatrix} \varepsilon & ig & 0 \\ -ig & \varepsilon & 0 \\ 0 & 0 & \varepsilon \end{pmatrix}. \qquad (1)$$

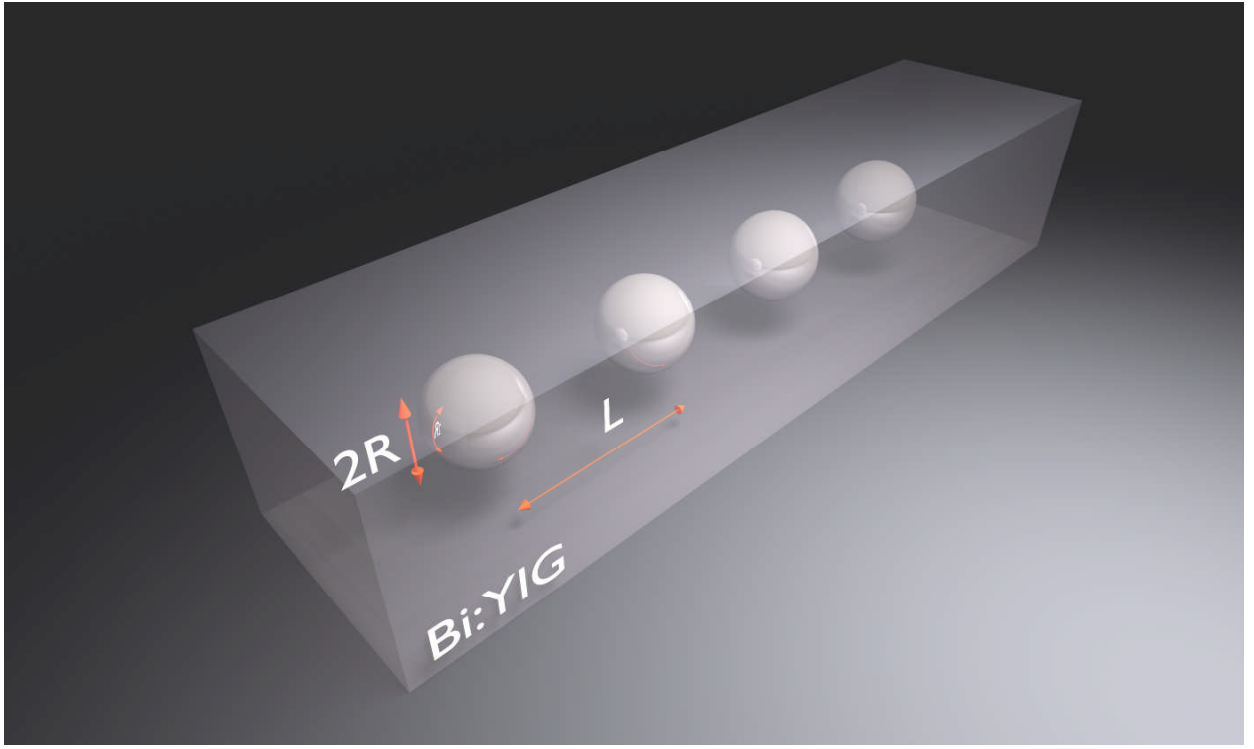

Fig. 1. Geometry of the problem: a periodic array of silver nanospheres each of radius $R$ with interparticle distance $L$ embedded into a MO medium.

We model the NPs as point dipoles. In order to find guided modes of the dipolar chain, one needs to know two essential characteristics: dipole polarizability of a spherical inclusion in a MO medium and the Green's function of an electric dipole in a MO medium that includes both far and near fields. However, far from the light cone (for wavevectors $k \gg \omega/c$) the dispersion of guided modes of the SDC is well described by nearest neighbor near field interactions,[29] which we use below. The dipole polarizability is calculated in the quasistatic approximation.

The electrostatic potential $\varphi$ in the surrounding MO host obeys the equation:

$$\mathrm{div}\mathbf{D} = \nabla \cdot (\hat{\varepsilon} \nabla \varphi) = 0. \qquad (2)$$

Non-diagonal terms in the permittivity tensor (1), $ig\partial_{xy}\varphi$ and $-ig\partial_{yx}\varphi$, cancel out and Eq. (2) transforms into the Laplace equation:

$$\varepsilon\Delta\varphi = 0. \tag{3}$$

This allows us to conclude that the near field of a point electric dipole **d** in an MO medium is simply the near field of an electric dipole embedded in an isotropic host medium with permittivity $\varepsilon$:

$$\mathbf{E}_{near}(\mathbf{r}) = -\frac{\mathbf{d}}{\varepsilon r^3} + 3\frac{(\mathbf{d}\cdot\mathbf{n})\mathbf{n}}{\varepsilon r^3}. \tag{4}$$

Now, let us find the polarizability dyadic of an isotropic sphere with permittivity $\varepsilon_{int}$ embedded into a MO medium. The sphere is subjected to an external oscillating homogenous field $\mathbf{E}e^{-i\omega t}$. Below we omit harmonic time dependence of the electric field $e^{-i\omega t}$. The electrostatic potential outside the sphere can be represented as

$$\varphi_{ext} = -\mathbf{E}\mathbf{r} + (\hat{A}\mathbf{E})\mathbf{r}/r^3, \tag{5}$$

where $\hat{A}$ is an unknown tensor which relates the applied electric field **E** and the induced dipole moment of the sphere **d**. The electric field in the surrounding medium is given by

$$\mathbf{E}_{ext} = -\nabla\varphi_{ext} = \mathbf{E} - \hat{A}\mathbf{E}/r^3 + 3(\hat{A}\mathbf{E}\cdot\mathbf{n})\mathbf{n}/r^3, \tag{6}$$

where $\mathbf{E}_{ext}\times\mathbf{n} = \mathbf{E}\times\mathbf{n} - \hat{A}\mathbf{E}\times\mathbf{n}/r^3$. The normal component of the displacement **D**, which is needed in the following calculations, can be expressed through the incident field **E** as

$$\mathbf{D}_{ext}\cdot\mathbf{n} = \hat{\varepsilon}\mathbf{E}_{ext}\cdot\mathbf{n} = \varepsilon\left(\mathbf{E}\cdot\mathbf{n} + 2\hat{A}\mathbf{E}\cdot\mathbf{n}/r^3\right) + \hat{G}\left(\mathbf{E}\cdot\mathbf{n} - \hat{A}\mathbf{E}\cdot\mathbf{n}/r^3\right), \tag{7}$$

where $\hat{G}$ is the non-diagonal part of the permittivity tensor $\hat{\varepsilon}$. The electric field inside the spherical inclusion can be written in the form

$$\mathbf{E}_{int} = \hat{B}\mathbf{E}, \tag{8}$$

where $\hat{B}$ is an as yet undetermined tensor. Using electromagnetic boundary conditions for Maxwell equations we obtain the system of equations:

$$\begin{cases} \hat{I} - \hat{A}/R^3 = \hat{B}, \\ \varepsilon(\hat{I} + 2\hat{A}/R^3) + \hat{G}(\hat{I} - \hat{A}/R^3) = \varepsilon_{int}\hat{B}, \end{cases} \tag{9}$$

which allows us to determine the tensor $\hat{A}$:

$$\hat{A} = R^3\left(\varepsilon_{int}\hat{I} - \varepsilon\hat{I} - \hat{G}\right)\left(\varepsilon_{int}\hat{I} + 2\varepsilon\hat{I} - \hat{G}\right)^{-1}. \tag{10}$$

Dipole polarizability relates the incident field and the induced dipole moment of a NP as $\mathbf{d} = \hat{\alpha}\mathbf{E}$. Comparing Eqs. (4) and (6) we conclude that $\hat{\alpha} = \varepsilon\hat{A}$. The elements of the dipole polarizability are given by:

$$\alpha_{xx} = \alpha_{yy} = \varepsilon R^3 \frac{\varepsilon_{int}^2 + \varepsilon_{int}\varepsilon - 2\varepsilon^2 - g^2}{\varepsilon_{int}^2 + 4\varepsilon_{int}\varepsilon + 4\varepsilon^2 - g^2},$$

$$\alpha_{xy} = -\alpha_{yx} = \varepsilon R^3 \frac{3i\varepsilon g}{\varepsilon_{int}^2 + 4\varepsilon_{int}\varepsilon + 4\varepsilon^2 - g^2}, \qquad (11)$$

$$\alpha_{zz} = \varepsilon R^3 \frac{\varepsilon_{int} - \varepsilon}{\varepsilon_{int} + 2\varepsilon}$$

and the polarizability dyadic is represented as

$$\hat{\alpha} = \begin{pmatrix} \alpha_{xx} & \alpha_{xy} & 0 \\ -\alpha_{xy} & \alpha_{xx} & 0 \\ 0 & 0 & \alpha_{zz} \end{pmatrix}. \qquad (12)$$

In the subspace $E_z = 0$, we find two eigenvalues of polarizability tensor $\hat{\alpha}$:

$$\alpha_+ = \alpha_{xx} + i\alpha_{xy} = \varepsilon R^3 \frac{\varepsilon^{int} - \varepsilon + g}{2\varepsilon + \varepsilon^{int} + g}, \quad \alpha_- = \alpha_{xx} - i\alpha_{xy} = \varepsilon R^3 \frac{\varepsilon^{int} - \varepsilon - g}{2\varepsilon + \varepsilon^{int} - g}. \qquad (13)$$

so that $\hat{\alpha}\mathbf{E}_+ = \alpha_+\mathbf{E}_+$ and $\hat{\alpha}\mathbf{E}_- = \alpha_-\mathbf{E}_-$, where $\mathbf{E}_+ = (1, i, 0)^T$ and $\mathbf{E}_- = (1, -i, 0)^T$ are the two eigenvectors of the polarizability tensor with $E_z = 0$. These two eigenvectors correspond to circularly polarized incident light. The third eigenvalue is equal to $\alpha_{zz}$ and is of no interest here because it corresponds to the electric field polarized along the magnetization vector for which no Faraday rotation can be observed.

### III. GUIDED MODES OF A MAGNETO-OPTICAL SUBDIFFRACTION CHAIN

Having calculated all the quantities in Section 2 one may find guided modes of a MO SDC. We seek guided solutions in the form $\mathbf{d}_n \sim e^{ikz_n}$. As shown earlier,[27] SDCs can support guided modes with both transverse (T) or longitudinal (L) polarizations. Since we are interested in the Faraday rotation which is naturally observed for T polarization, we limit ourselves to analyzing this particular polarization. Polarizations of the electric field and the NP dipole moment for the T-polarization are $\mathbf{E} = (E_x, E_y, 0)$ and $\mathbf{d} = (d_x, d_y, 0)$, respectively. For the transverse polarization of the electric field $(\mathbf{d}\cdot\mathbf{r}_n) = 0$, where $\mathbf{r}_n$ points along the chain, so that only a single term remains in Eq. (4).

Consistent analysis of the problem requires calculation of the total local electric field acting on a NP. This field is the sum of fields due to *all other dipoles in the chain*. However, as noted above, in the guided wave region far from the light cone, where $k \gg \omega/c$, the tight-binding approximation gives reasonable results which may be utilized to further analyze the problem.

In the tight-binding approximation, the electric field applied to the *n*-th point dipole is the sum of the fields created by the two neighboring particles:

$$\mathbf{E}_n = -\frac{\mathbf{d}_{n-1}}{\varepsilon L^3} - \frac{\mathbf{d}_{n+1}}{\varepsilon L^3} = -\frac{1}{\varepsilon L^3}\left(e^{-ikL}\hat{\alpha}\mathbf{E}_n + e^{ikL}\hat{\alpha}\mathbf{E}_n\right). \tag{14}$$

Equation (14) represents an eigenvalue problem for the polarizability tensor (12) of a metallic inclusion. Substituting eigenvalues (11) into Eqs. (13) for the two different eigenvectors $\mathbf{E}_+$ and $\mathbf{E}_-$, we obtain a pair of dispersion relations:

$$\mathbf{E}_\pm = -\frac{1}{\varepsilon L^3}\left(e^{-ikL} + e^{ikL}\right)\alpha_\pm \mathbf{E}_\pm = -\frac{2\cos kL}{\varepsilon L^3}\alpha_\pm \mathbf{E}_\pm. \tag{15}$$

Equation (15) leads to explicit expressions for the guided mode wavevectors:

$$k_\pm = \frac{1}{L}\cos^{-1}\left(\frac{-\varepsilon L^3}{2\alpha_\pm}\right). \tag{16}$$

According to Eq. (16) two guided modes, which we refer to as '+' and '–', have circular polarizations of the electric field. Taking into account different values of the Bloch wavevectors and circular polarization of the electric field, one may anticipate that propagation of an excitation along the chain is accompanied by the Faraday rotation.

Firstly, we present the results of numerical calculation for the case of lossless silver NPs embedded into a lossless MO medium for *R* = 10 nm and *L* = 4*R*. The two dispersion curves of guided modes of a 1D chain of silver NPs are shown in the Fig. 2(a). Experimental data from Refs. [38] and [39] was adopted to approximate permittivity of silver and Bi:YIG, respectively. One can see that in a relatively narrow frequency region between 2.2 eV and 2.4 eV, the plasmon resonance gives rise to the propagation band of an array of NPs. The two dispersion curves are split due to MO activity of the host medium, which manifests itself in different polarizabilities $\alpha_+$ and $\alpha_-$.

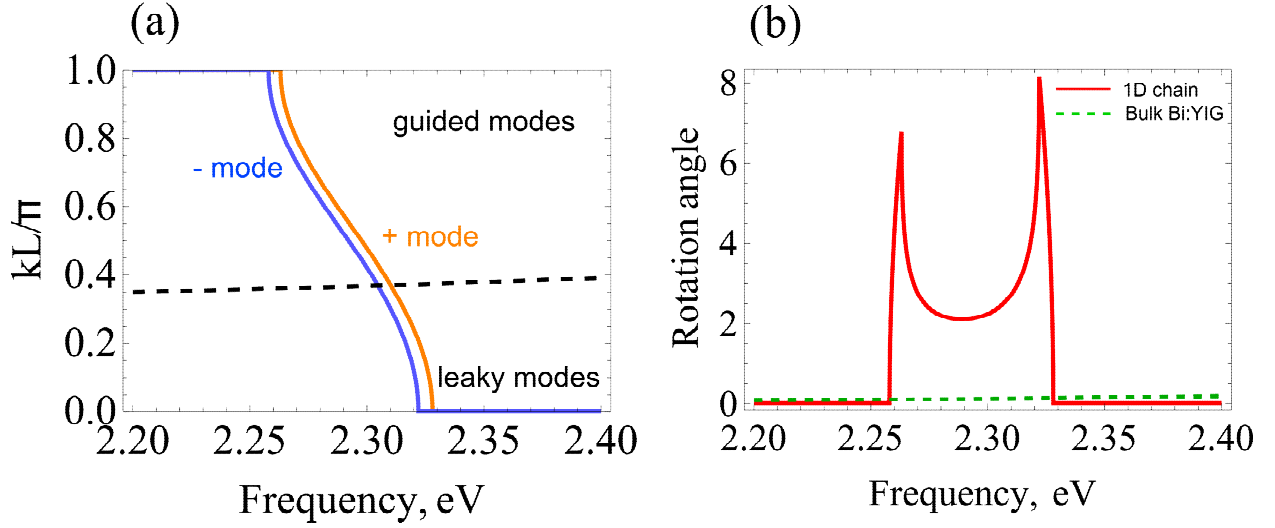

Fig. 2. (Color online) (a) Dispersion curves of the 1D chain of lossless silver NPs embedded into Bi:YIG. The dashed line indicates the light cone with $k = \omega/c$. The region below this line corresponds to leaky modes. (b) Estimation for the Faraday rotation per 1μm for the 1D chain of lossless silver NPs (solid curve) and for bulk Bi:YIG (dashed curve).

As mentioned above, we search for guided solutions only in the region for which $|\mathrm{Re}(k)| > q$, where $q = \mathrm{Re}(n_{Bi:YIG}\omega/c)$ is the wavevector of a plane wave travelling in the host medium. Excitations with $\mathrm{Re}(k) > q$ can propagate along the chain without radiative losses, while modes with $\mathrm{Re}(k) < q$ experience strong radiative decay and are not considered in the paper.

The Faraday rotation of the polarization by a MO chain per unit length can be estimated as $\theta = k_+ - k_-$. Indeed, when a chain is excited at its input by a linearly polarized electric field $E_{input} = (1,0,0)^T$, which is the most common situation, this polarization is decomposed into two eigenmodes of a chain with equal amplitudes:

$$E_{input} = \frac{1}{2}\left[(1,i,0)^T e^{ik_+ z} + (1,-i,0)^T e^{ik_- z}\right]\bigg|_{z=0}. \qquad (17)$$

This is similar to the case of a uniform MO medium, for which the polarization rotation is also given by the above expression. To obtain an estimate for $\theta$, we calculate rotation angle per chain length of 1μm, which is given by $\theta = 10^{-6}\,\mu\mathrm{m}\cdot(k_+ - k_-)/2$. This estimation is plotted in Fig. 2(b).

As we can see from Fig. 2(b), the Faraday rotation is dramatically increased compared to bulk Bi:YIG. Notably, the Faraday rotation increases even more in the vicinity of the band gap. This behavior can be understood from Fig. 2(a). Indeed, near the band edge of one mode, when the other is still propagating, the group velocity, $v_g = \partial \omega / \partial k$, of this mode rapidly drops to zero, which results in a larger distance (and larger wavevectors difference) between the two modes. As a consequence, the Faraday rotation angle, $\theta = k_+ - k_-$, increases as well.

The remarkable property of dispersion law (16) is its *scaling behavior*. One may fix the ratio $R/L$ and tend both NP's radius, $R$, and interparticle distance, $L$, to zero. Then, wavevectors of '+' and '−' guided modes proportionally increase as $1/L$. The Faraday rotation angle, $\theta$, increases as well. In fact, the factor, which limits such a scaling behavior, is applicability of the concept of permeability. When a NP is small enough ($<5\,\text{nm}$), the description of the NP in terms of permittivity tensor (1) is not applicable and its electromagnetic response no longer can be described by polarizability $\hat{\alpha}$ (13)[1].

Dispersions of guided modes and the Faraday rotation in the case of lossy NPs and the host medium are presented in Fig. 3. Overall, the propagation band becomes less pronounced. The propagation length of the guided mode can be estimated as $l_{prop} = |\text{Im}\, k|^{-1}$. As seen from Fig. 3(a), the propagation distance is of the order of the interparticle distance $L$. This means that the wave is absorbed before it reaches the end of the chain. Ohmic losses also have strong impact on the Faraday rotation angle: the maximum value of the Faraday rotation in chain of lossy particles is several times lower than that of lossless particles.

Note that the dispersion curves obtained from Eqs. (16) represent backward waves: the group velocity and the wavevector have opposite directions (see Fig. 3). Indeed, the waves are travelling in the direction of the group velocity, which is chosen as positive. Obviously, propagating in this direction the wave loses energy. We choose this way of plotting to ensure that the group velocities and imaginary parts of wavevectors $k_\pm$ are positive, whereas the real part of wavevectors is negative.

Results presented in this section allow us to conclude that for Faraday rotation enhancement, it is critical to have low value of Ohmic losses. A practical way of achieving this is compensation of the effect of losses. In the next section, we show how Ohmic loss can be minimized in gain-assisted chains formed by composite core-shell NPs.

---

[1] We are appreciative to Antonio García-Martín who has directed our attention to this fact.

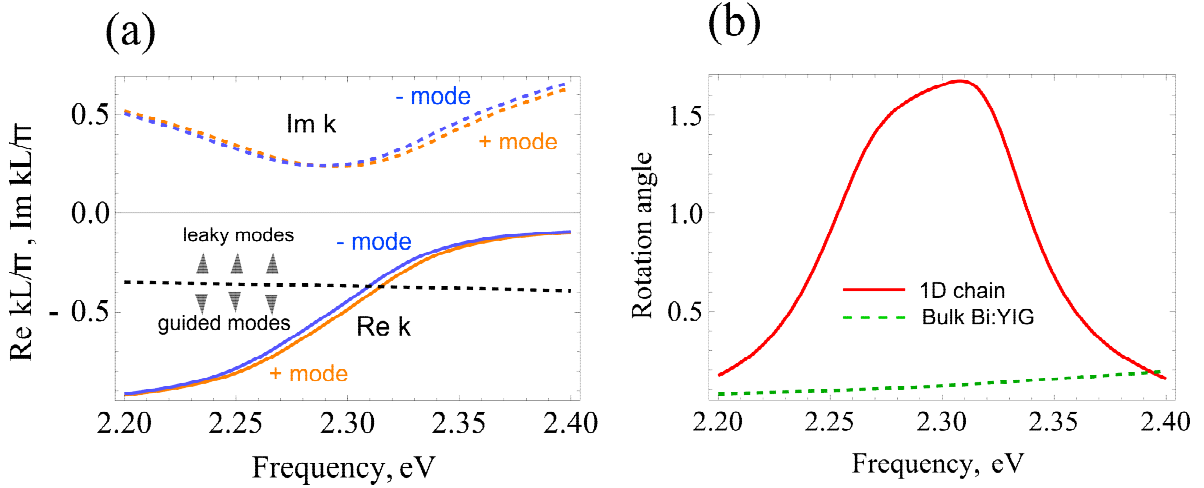

Fig. 3. (Color online) (a) Dispersion of guided modes of lossy MO chain. Solid curves: real parts of the Bloch wavevectors $\mathrm{Re}\, k_\pm$, dashed curves: imaginary parts of wavevectors $\mathrm{Im}\, k_\pm$. Black dashed line separate the guided and leaky regions. (b) Estimates for the Faraday rotation angles per 1 μm for lossy MO chain (solid curve) and for bulk Bi:YIG (dashed curve).

## IV. GAIN-ASSISTED SUBDIFFRACTION MAGNETO-OPTICAL CHAIN

The problem of strong attenuation in SDCs due to Ohmic damping has been addressed in the literature.[40-42] In Ref. [40] amplification is provided by gain in the host medium, into which the plasmonic chain is embedded, while in Ref. [42] authors consider a chain formed by composite core-shell NPs in which gain is provided by active cores. In Ref. [41] both approaches are analyzed and compared. We follow the second approach and consider a SDC made of composite gain-assisted NPs. A single composite NP is schematically depicted in Fig. 4. It consists of the silver spherical core of radius $r_1$ and the gain shell of radius $r_2$. The layout of a chain is the same as in the previous section.

Combination of a gain medium and a MO plasmonic resonator leads to the formation of the MO spaser[43] (for a review of spasers see Ref. [44]) Above the threshold, it has two lasing modes with circular polarizations of dipole moments and different lasing frequencies. Here we are more concerned with the below-threshold regime of the MO spaser, when gain is insufficient to support spasing.

The electromagnetic response of the gain medium below the lasing threshold can be described with an effective permittivity, which is deduced from the Maxwell-Bloch equations:[45-47]

$$\varepsilon_{gain}(\omega) = \varepsilon_0 + D_0 \frac{\omega_0}{\omega} \frac{-i + \frac{\omega^2 - \omega_0^2}{2\omega\Gamma}}{1 + \left(\frac{\omega^2 - \omega_0^2}{2\omega\Gamma}\right)^2}, \tag{17}$$

where $\Gamma$ is the emission linewidth of the gain medium, $\omega_0$ is the emission frequency and $D_0$ is a dimensionless characteristic of the pump rate, which is proportional to the population inversion of the quantum emitter without an incident field. For organic dye molecules, which may serve as the gain medium, the typical value of parameter $\Gamma$ is $\Gamma \sim 10^{-14}$ s.[48] Background permittivity of the gain medium is set to $\varepsilon_0 = 4$ and its emission frequency is tuned to the narrow propagation band of SDC. The radii of core-shell NP are $r_1 = 10$ nm and $r_2 = 1.2 r_1$, respectively. For the parameters specified above, the propagation band of a SDC arises in the frequency region between 2.4 eV and 2.45 eV, therefore, the emission frequency is set to $\omega_0 = 2.42$ eV.

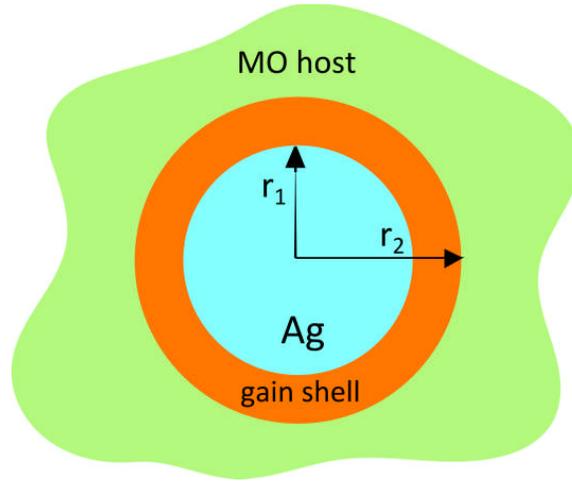

Fig. 4. A schematic drawing of a gain-assisted core-shell NP constituting gain-assisted MO chain.

Following the same procedure as in Section 3, we find the dispersion law of guided modes in the tight-binding approximation:

$$k_\pm = \frac{1}{L} \cos^{-1}\left(\frac{-\varepsilon L^3}{2\alpha_\pm^{core-shell}}\right), \tag{18}$$

where $\alpha_\pm^{core-shell}$ are eigenvalues of the dipole polarizability tensor of the composite core-shell NP placed in the MO medium. This tensor is found in the same fashion as for the case of a simple spherical NP in a MO medium. Expressions for these two eigenvalues can be found in the Appendix.

Basically, in most applications a non-zero output signal and dipole moments of NPs are only generated by a non-zero input signal. At the same time, in gain systems, when the pump rate exceeds the threshold, an electrodynamic system of dipole scatterers may have lasing states, at which the system possesses non-zero dipole moments without any incident radiation. Such a regime of operation is inappropriate for the purpose of the polarization control. Below we assume that the pump rate is smaller than the lasing threshold.

The threshold appears as a pole of dipole polarizability, namely, as a pole of one of its eigenvalues. Since dyadic $\hat{\alpha}^{core-shell}$ has three different eigenvalues corresponding to different polarizations of incident light, each eigenvalue has its own threshold and corresponds to the unique configuration of the lasing mode. To prevent lasing one should choose pump rate $D_0$ below the lowest threshold $D_{min}$.

Situation changes for a *system of interacting NPs with gain*. Due to the effect of multiple scattering, in such a system, thresholds for each mode changes. In this paper we do not establish the general condition for lasing in a system of interacting dipoles.[49] Instead, we use less general but more strict condition which guarantees that no lasing occurs in a periodic chain.

Scattering behavior of a composite NP is described by dipole polarizability $\hat{\alpha}^{core-shell}$. This dyadic may be diagonalized so that scattering may be described by its three elements in the diagonal basis: $\alpha_+^{core-shell}$, $\alpha_-^{core-shell}$ and $\alpha_{zz}^{core-shell}$. When imaginary parts of this quantities are all positive,

$$\text{Im}\,\alpha_+ > 0, \ \text{Im}\,\alpha_- > 0, \ \text{Im}\,\alpha_{zz} > 0, \tag{19}$$

scattering of incident light for any possible polarization is accompanied by dissipation, since the dissipated power is given by

$$W = \frac{\omega}{2}\left(\text{Im}\,\alpha_+ |E_+|^2 + \text{Im}\,\alpha_- |E_-|^2 + \text{Im}\,\alpha_{zz} |E_z|^2\right). \tag{20}$$

On the other hand, for a system of dissipative dipole scatterers, a lasing mode cannot arise regardless of the value of gain in each NP. Indeed, under this condition, the total dissipated power in the system is always positive and lasing mode cannot exist.

Relying on this simple argument, we limit the range of the pump rate to such values, for which inequalities (19) hold. The maximum value of the pump rate at which these three conditions are satisfied, we denote by $D_{diss}$. For values of the pump rate such that $D_0 < D_{diss}$, the core-shell NP is strictly dissipative. Although we cannot analytically establish the value of $D_{diss}$ due to the very complex form of the expression for $\hat{\alpha}^{core-shell}$, numerical calculations for the parameters of NP introduced above indicate that $0.3 < D_{diss}$. From Fig. 5 one can see that for $D_0 = 0.3$, all three eigenvalues demonstrate dissipative behavior. For any $D_0 < 0.3 < D_{diss}$, all three eigenvalues have positive imaginary parts.

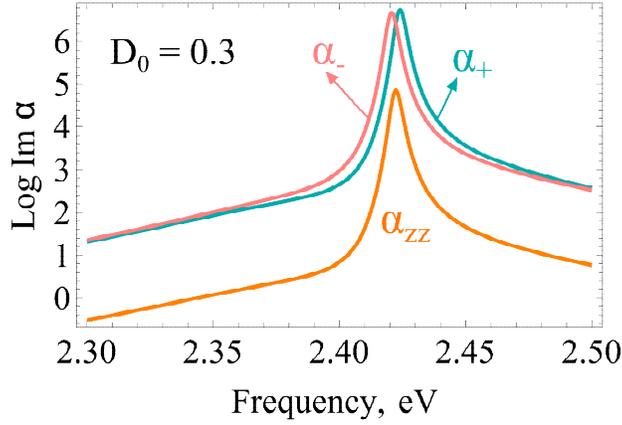

Fig. 5. Eigenvalues of polarizability dyad $\hat{\alpha}^{core-shell}$ for the pump rate $D_0 = 0.3$. Since imaginary parts of all three eigenvalues are positive, the system is dissipative.

In Fig. 6, we plot dispersion of guided modes of gain-assisted MO chain for zero pump rate, Figs. 5(a) and (b), and for the optimal value of the pump rate $D_0 = 0.3 < D_{diss}$, Figs. 6(c) and (d). The interparticle distance is $L = 4r_2$. Overall, patterns of dispersion curves and the Faraday rotation in a passive configuration with zero pump rate follow that of a passive chain of silver NPs [see Fig. 3(a) and (b)]. Real and imaginary parts of the wavevector have similar values [Figs. 6(a) and (c)] and thus, the propagation distance of the guided mode is of the order of Bloch wavelength $2\pi / \mathrm{Re}\, k$.

As the pump rate, the propagation band of a MO chain becomes narrower and more pronounced [Fig. 6(c)]. Due to the gain layer at the surface of silver NP, the plasmon resonance frequency slightly detunes from that of a bare silver NP immersed into a MO medium. This results in a slightly different frequency region of the propagation band. Fig. 6(c) also indicates that due to gain, the propagation length of the guided mode of MO chain increases (the imaginary part of the complex wavevector decrease). This increase is substantial: when gain is on, the propagation distance is approximately tenfold greater compared to the passive case.

At the same time, the Faraday rotation also increases with the rise of the pump rate $D_0$. For $D_0 = 0.3$, the Faraday rotation is comparable or greater than that in a lossless SDC [Figs. 6(b) and (d)]. Interestingly, similar to a lossless SDC, in a gain-assisted chain, the Faraday rotation curve has a fine structure within the narrow propagation band: the Faraday rotation increases near band edges [see Fig. 6(d)].

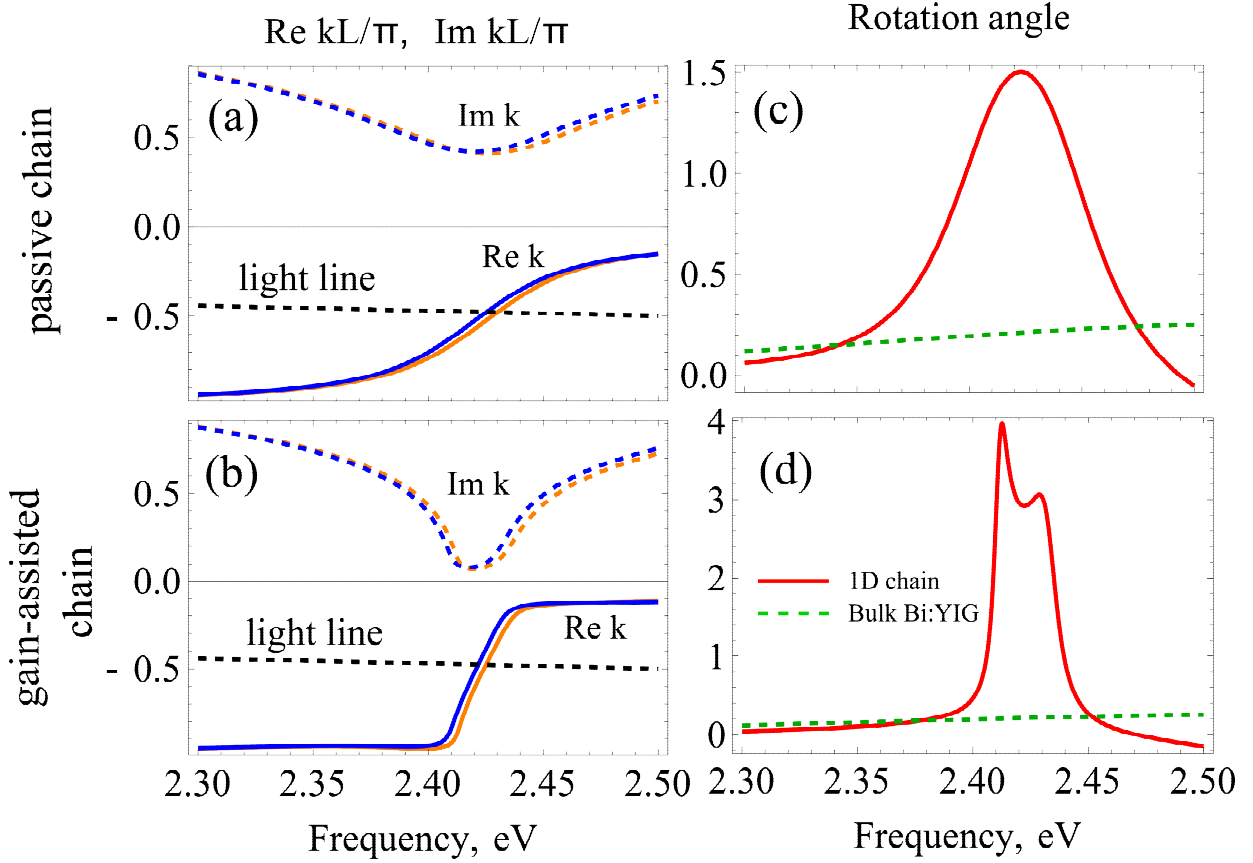

Fig. 6. Left panel: dispersion of guided modes of composite core-shell NPs embedded into YIG. Right panel: the Faraday rotation along the chain per $1\,\mu m$. Top: passive SDC with zero the pump rate. Bottom: gain-assisted SDC with $D_0 = 0.3$.

Figure 6 shows the Faraday rotation as a function of mode frequency and the pump rate $D_0$. As the pump rate increases, the rotation angle also gradually increases within the propagation band. When the pump rate reaches the value close to $D_0 = 0.3$, the fine structure with two distinct peaks becomes visible. These peaks as argued previously correspond to the propagation band edge, at which the group velocity of one of the modes drops to zero.

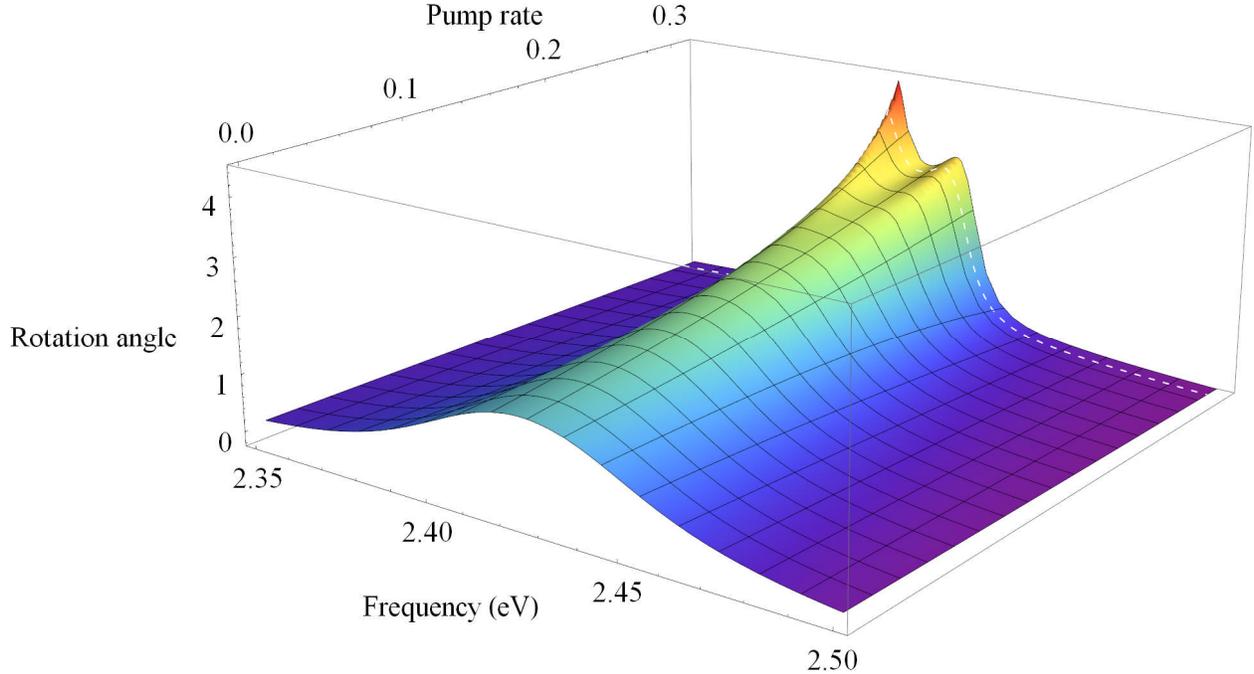

Fig. 7. The Faraday rotation per $1\,\mu m$ as a function of the mode frequency and the pump rate, $D_0$. The dashed line indicates frequency-dependent rotation at pump rate of $D_0 = 0.3$ used in Fig. 6(d).

## V.  CONCLUSION

We have carried out a theoretical study of the enhancement of the Faraday rotation in a 1D periodic chain of plasmonic NPs embedded into a MO medium. Using parameters for silver NPs and Bi-substituted yttrium iron garnet MO we show that in this system, the angle of rotation of the direction of the NP dipole polarization increases by orders of magnitude compared to a bulk MO medium. Ohmic losses, however, substantially suppress this effect. We propose a way of compensating for these losses by covering NPs with gain layers so that each NP turns into a MO spaser. The pump rate should be smaller than the spasing threshold. When the pumping frequency is tuned to the surface plasmon resonance, the imaginary part of the wavevector of the guided mode will vanish and the Faraday rotation will be resonantly enhanced by two orders of magnitude.

## ACKNOWLEDGEMENTS

This work was supported by RFBR grants No. 13-07-92660, 12-02-01093-a, 13-02-00407, by the Dynasty Foundation, and by the NSF under Grant No. DMR-1312707.

## APPENDIX: POLARIZABILITY OF A CORE-SHELL NANOPARTICLE IN A MAGNETO-OPTICAL MEDIUM

Consider a core-shell NP in a MO medium with inner and outer radii $r_1$ and $r_2$, respectively. To find polarizability dyadic of a composite core-shell NP embedded into a MO host in the quasi-static approximation, we write the electric field in the outer region (the MO host medium) as

$$\mathbf{E}_{ext}(\mathbf{r}) = \mathbf{E} - \frac{\hat{A}\mathbf{E}}{r^3} + 3\frac{(\hat{A}\mathbf{E}\cdot\mathbf{n})}{r^3}\mathbf{n}. \tag{A1}$$

The electric field in the shell can be represented as a sum of a homogenous field and point dipole-like contribution:

$$\mathbf{E}_{shell}(\mathbf{r}) = \hat{B}\mathbf{E} - \frac{\hat{C}\mathbf{E}}{r^3} + 3\frac{(\hat{C}\mathbf{E}\cdot\mathbf{n})}{r^3}\mathbf{n}. \tag{A2}$$

In Eqs. (A1) and (A2), tensors $\hat{A}$, $\hat{B}$, and $\hat{C}$ are unknown. The electric field in the core is represented only by the homogenous term, related to the incident field by another unknown tensor:

$$\mathbf{E}_{core}(\mathbf{r}) = \hat{F}\mathbf{E}. \tag{A3}$$

In the outer region, the near field of the NP has the same structure as that of a solid metallic NP in a MO medium [see Eq. (6)], therefore dyadic polarizability is expressed as $\hat{\alpha}^{core-shell} = \varepsilon\hat{A}$ and our goal is to find this tensor from the Maxwell's boundary conditions.

Again, normal and tangential components of electric field and displacement, respectively, in outer, shell, and core regions are expressed as:

$$\begin{aligned}
\mathbf{E}_{ext} \times \mathbf{n} &= \mathbf{E}\times\mathbf{n} - \hat{A}\mathbf{E}\times\mathbf{n}/r^3, \\
\mathbf{D}_{ext} \cdot \mathbf{n} &= \varepsilon\left(\mathbf{E}\cdot\mathbf{n} + 2\hat{A}\mathbf{E}\cdot\mathbf{n}/r^3\right) + \hat{G}\left(\mathbf{E}\cdot\mathbf{n} - \hat{A}\mathbf{E}\cdot\mathbf{n}/r^3\right), \\
\mathbf{E}_{shell} \times \mathbf{n} &= \hat{B}\mathbf{E}\times\mathbf{n} - \hat{C}\mathbf{E}\times\mathbf{n}/r^3, \\
\mathbf{D}_{shell} \cdot \mathbf{n} &= \varepsilon_{gain}\left(\hat{B}\mathbf{E}\cdot\mathbf{n} + 2\hat{C}\mathbf{E}\cdot\mathbf{n}/r^3\right), \\
\mathbf{E}_{core} \times \mathbf{n} &= \hat{F}\mathbf{E}\times\mathbf{n}, \\
\mathbf{D}_{core} \cdot \mathbf{n} &= \varepsilon_{core}\hat{F}\mathbf{E}\cdot\mathbf{n}.
\end{aligned} \tag{A4}$$

Substituting these expressions into the boundary conditions we arrive at the system of equations determining all introduced tensors:

$$\begin{cases} \hat{I} - \hat{A}/r_2^3 = \hat{B} - \hat{C}/r_2^3, \\ \varepsilon(\hat{I} + 2\hat{A}/r_2^3) + \hat{G}(\hat{I} - \hat{A}/r_2^3) = \varepsilon_{shell}\left(\hat{B} + 2\hat{C}/r_2^3\right), \\ \hat{B} - \hat{C}/r_1^3 = \hat{F}, \\ \varepsilon_{shell}(\hat{B} + 2\hat{C}/r_1^3) = \varepsilon_{core}\hat{F}. \end{cases} \quad (A5)$$

Finally, from this system we find tensor $\hat{A}$ and dipole polarizability $\hat{\alpha}^{core-shell}$. Explicit expressions for polarizability eigenvalues are:

$$\alpha_+ = r_2^3 \varepsilon \left(1 + \frac{3\varepsilon\left(\varepsilon_{core}\left(r_1^3 - r_2^3\right) + \varepsilon_{shell}\left(r_1^3 + 2r_2^3\right)\right)}{\varepsilon_{core}\left((2\varepsilon+g)\left(r_1^3 - r_2^3\right) - \varepsilon_{shell}\left(2r_1^3 + r_2^3\right)\right) + \varepsilon_{shell}\left(-(2\varepsilon+g)\left(r_1^3 + 2r_2^3\right) + 2\varepsilon_{shell}\left(r_1^3 - r_2^3\right)\right)}\right),$$

$$\alpha_- = r_2^3 \varepsilon \left(1 + \frac{3\varepsilon\left(\varepsilon_{core}\left(r_1^3 - r_2^3\right) + \varepsilon_{shell}\left(r_1^3 + 2r_2^3\right)\right)}{\varepsilon_{core}\left((2\varepsilon-g)\left(r_1^3 - r_2^3\right) - \varepsilon_{shell}\left(2r_1^3 + r_2^3\right)\right) + \varepsilon_{shell}\left(-(2\varepsilon-g)\left(r_1^3 + 2r_2^3\right) + 2\varepsilon_{shell}\left(r_1^3 - r_2^3\right)\right)}\right),$$

$$\alpha_{zz} = r_2^3 \varepsilon \left(1 - \frac{3\varepsilon\left(\varepsilon_{core}\left(r_1^3 - r_2^3\right) + \varepsilon_{shell}\left(r_1^3 + 2r_2^3\right)\right)}{2\varepsilon_{shell}\left(\varepsilon\left(r_1^3 + 2r_2^3\right) + \varepsilon_{shell}\left(-r_1^3 + r_2^3\right)\right) + \varepsilon_{core}\left(2\varepsilon\left(-r_1^3 + 2r_2^3\right) + \varepsilon_{shell}\left(2r_1^3 + r_2^3\right)\right)}\right),$$

(A6)

where $\alpha_\pm$ correspond to polarizations of incident light $\mathbf{E} = (1, \pm i, 0)^T$ and $\alpha_{zz}$ corresponds to incident light polarized along the magnetization vector of the MO host medium.